\documentclass[twocolumn,showpacs,preprintnumbers,amsmath,amssymb]{revtex4}
\usepackage{graphicx}
\usepackage{dcolumn}
\usepackage{bm}

\def\miset {E_T\!\!\!\!\!\!\!/ \;\;\,}

\def\beq{\begin{equation}}
\def\eeq{\end{equation}}

\newcommand{\lsim}
{{\;\raise0.3ex\hbox{$<$\kern-0.75em\raise-1.1ex\hbox{$\sim$}}\;}}
\newcommand{\gsim}
{{\;\raise0.3ex\hbox{$>$\kern-0.75em\raise-1.1ex\hbox{$\sim$}}\;}}
 
\begin{document}
\preprint{MRI-P-031102}
\preprint{hep-ph/0311347}


\title {Gauge boson fusion as a probe 
of inverted hierarchies in supersymmetry}

\author{Partha Konar}
 \email{konar@theory.tifr.res.in}
 \altaffiliation[\\ Present address: ]
 {Department of Theoretical Physics,
  Tata Institute of Fundamental Research,
  Homi Bhabha Road,
  Mumbai-400005,
  India}
\author{Biswarup Mukhopadhyaya}
 \email{biswarup@mri.ernet.in}

\affiliation{ Harish-Chandra  Research Institute,\\
 Chhatnag Road, Jhusi, Allahabad - 211 019, India }


\begin{abstract}
Supersymmetric scenarios with inverted mass hierarchy can be 
hard to observe at a hadron collider, particularly
in the non-strongly interacting sector. We show how the production
of stau-pairs via gauge boson fusion, along with hard jets in the
high rapidity region, can be instrumental in uncovering the
signatures of such scenarios. We demonstrate this both in a model-independent
way and with reference to some specific, well-motivated models. 
\end{abstract}

\pacs{12.60.Jv, 13.35.Dx, 13.85.Rm}

\maketitle

\noindent
Supersymmetry (SUSY) is perhaps the most frequently discussed new physics
\cite{susyrev} that is expected to exist around the TeV scale. Such a scale 
is  attributed to SUSY because that is how it can aspire to lend naturalness 
to the electroweak theory. The fact remains, however, that neither have 
we found any experimental signal of SUSY yet, nor is there an unambiguous  
guideline on the superparticle spectrum or the mechanism of SUSY breaking 
which is so essential to make the theory realistic. Still, the very 
necessity of solving the  naturalness problem raises hopes of 
discovering superparticles at TeV scale
colliders such as the Large Hadron Collider (LHC).

On the other hand, most SUSY  theories are beset with the flavour 
problem \cite{flav},
which essentially means the danger of having unacceptable enhancement of
flavour-changing neutral current (FCNC) processes. One way to avoid this
difficulty is to have the mass scale of superparticles raised to several,
often tens of, TeV. However, this largely defeats the purpose of
introducing SUSY to solve the naturalness problem. A possible way out lies 
in theories which have the third family of scalar fermions light,
against the backdrop of a heavy matter sector in the first two families.
Such `inverted hierarchy' has been achieved in a number of theoretical
frameworks. This can be done, for example, through  \\
(a) SUSY breaking induced by modular and dilaton fields\cite{brig}, with the
modular weight being different for different families, thus leading to
a lower scalar mass for the third family at high scale itself.\\
(b) Introducing some additional (anomalous) $U(1)$ symmetry, with
family-dependent $U(1)$ charges, thus allowing the consequent D-terms to 
lower the third family scalar masses \cite{u1}.\\
(c) Non-universal boundary conditions at the Grand Unification (GUT)
scale \cite{gutcond}, with other high-scale boundary 
conditions suitably adjusted,
and by demanding Yukawa coupling unification (thus allowing third
family scalars to be affected by large Yukawa couplings as they run).\\
(d) Arranging SUSY parameters in such a way that the third family masses
have fixed points below a TeV \cite{fixp}.

Diverse as the phenomenological consequences of the above cases may be,
all of them pose a serious question: how can the non-strongly 
interacting sfermion sector be revealed in experiments? This is because
sleptons are usually expected to be seen in the Drell-Yan
channel, where the production rates fall to rather low values for 
$m_{\tilde{l}}~~\simeq~~250 - 300$ GeV \cite{tata_slep}. 
Stau's ($\tilde{\tau}$)
in inverted hierarchy scenarios, even if still
marginally accessible in the Drell-Yan channel, have their signals 
further suppressed because of the complications involved in identifying
tau's. The resulting difficulties are again twofold. First of all,
if charginos and neutralinos, too, are almost as heavy as the staus,
their detectability (in hadronically quiet channels such as 
trileptons \cite{trilep}) falls below the threshold of detection at the LHC.
Alternatively, if charginos and neutralinos are relatively light, then 
they may be detected, while we have little information on 
the SUSY particle spectrum, and cannot even confirm an inverted hierarchy.

Here we suggest a new search channel for SUSY scenarios
with inverted hierarchy, using gauge boson fusion at the LHC to
produce stau-pairs. We show that this not only makes the stau signals 
relatively background-free, but also enhances the mass reach 
for the stau's, thus opening a gateway to scenarios of this kind. 

Gauge boson fusion has been found to be a useful channel for exploring
the signals of a heavy Higgs boson \cite{dawson}. Subsequent studies have also
underlined its usefulness for an intermediate mass Higgs, 
especially for Higgs decay modes such as those into
$\tau\tau$, $\gamma\gamma$ or $b\bar{b}$, or for probing
couplings which can potentially distinguish a supersymmetric 
Higgs boson \cite{zeppenfeld}. The characteristic features of such events are 
two hard forward jets, usually peaking in the rapidity region 
$3 < |\eta| < 4$, with the lack of colour exchange between the jets 
preventing hadronic activity in the intervening rapidity gap \cite{bjorken}.
Tagging the forward jets reduces the backgrounds drastically.
Also, such events survive a central jet veto with a high
($\ge 80$\%) efficiency. It is because of all this that the facility of
forward jet tagging is going to be an integral part of detector 
design at the LHC.

It has been shown in a series of recent studies that gauge 
boson fusion 
can be also very helpful in unraveling the signatures of physics beyond 
the standard model. This has been demonstrated mostly in the 
context of supersymmetric theories, for example, ones with invisible 
charginos and neutralinos \cite{konar} or sleptons \cite{sl_vbf}
with masses on the heavier side. 
Gauge boson fusion lends visibility to the latter situation when the 
conventional Drell-Yan signal becomes too small to be detectable.  
In the same way, one can also see signals of the stau 
when the latter is the only non-strongly interacting supersymmetric 
particle.

The signal we are suggesting comes from
\begin{eqnarray}
pp \longrightarrow j_f j_f \tilde{\tau}\tilde{\tau} \longrightarrow 
j_f j_f \tau \tau
+ \miset
\end{eqnarray}
 
\noindent
$j_f$ being a hard forward jet.
The missing transverse energy comes from the lightest neutralino
due to stau-decay. The $\tilde{\tau}$ (and $\tau$) decay products lie in the
rapidity gap between the forward jets, with no other colour activity
in that region.  In order to observe the $\tau$'s, we suggest the events
where one of them decays leptonically and the other, into the one-prong
hadronic channel. Therefore, the final state in this channel consists of 
$j_f j_f l j_{\tau} + \miset$,
$j_{\tau}$ being a low-multiplicity jet characteristic of $\tau$-decay.

In practice, however, there is a large region of the SUSY parameter 
space where the stau has a substantial branching ratio
for decay into the lighter chargino or the second lightest 
neutralino. This happens particularly when the stau mass is
well above that of the lightest SUSY particle (LSP).
In such cases, the loss of signal events due to branching fraction 
suppression may be partially offset by including events where the stau 
decays into a chargino and the latter, in the leptonic channel.
Such a possibility has been included in our calculation.

A large number of diagrams contribute to the above process. Stau-pair 
production in the desired form can take place through the fusion of
the W, the Z as well as the photon. All the production modes, namely,
 $\tilde{\tau}_1$- $\tilde{\tau}_1$,  $\tilde{\tau}_2$- $\tilde{\tau}_2$ 
and  $\tilde{\tau}_1$- $\tilde{\tau}_2$ are included in 
the general analysis. Gauge invariance requires one to include 
subprocesses other than those involving gauge boson fusion, although
they contribute very little when all the event selection criteria are
imposed. In addition to electroweak subprocesses, it is also
necessary to take into account the real emission corrections to Drell-Yan
production; being  strong processes, they have large rates, although 
the survival probability under a central jet veto is rather low.
We have used the survival probability to be 80\% (15\%) for electroweak
(QCD) subprocesses \cite{rainthes}.

\begin{table}[ht]
\caption{\label{table1} Signal and background cross sections surviving 
each type of cuts, for $M_{\tilde{\tau}_1} =400 ~GeV$, 
$M_{\tilde{\tau}_2} =430, ~GeV$ $\cos \theta_\tau = 0.9$ and 
$M_2 = 400 ~GeV$. Basic cuts are as specified in the text. The $t\bar{t}$ 
background includes $t\bar{t}+jets$.}
\begin{ruledtabular}
\begin{tabular}{lcccccc}
\emph{ }&
\emph{Signal}&
\multicolumn{5}{c}{\emph{Background (in fb)}}\\ 
&
\emph{(in fb) } & 
\emph{ $\tau \tau $ }&
\emph{ $W j $ } &  
\emph{ $W W $ } &
\emph{ $t \bar{t}$ } &
\emph{ Total}  \\

 \bf Basic cuts                 & 0.73 & 2.88 & 2.01 & 0.37 & 5.06 & 10.30 \\ 

+$M_{j_f j_f}> 1200 ~GeV$       & 0.57 & 1.37 & 0.63 & 0.25 & 1.14 & 3.41 \\ 

+$\miset > 100 ~GeV$            & 0.42 & 0.32 & 0.09 & 0.11 & 0.24 & 0.77 \\ 

+$M_{lj_{\tau}} > 60 ~GeV$      & 0.31 & 0.03 & 0.08 & 0.10 & 0.16 & 0.38 \\ 
\end{tabular}
\end{ruledtabular}
\end{table}

Our calculation is done in the helicity amplitude formalism, using the
subroutine HELAS \cite{helas}. All calculations corresponds to
the LHC energy ($\sqrt{s}=14$ TeV), and CTEQ4L structure 
functions\cite{cteq} have been used.
The following `basic cuts' are employed to
ensure the {\it bona fide} of the gauge boson fusion events:\\
(a) Two forward jets in opposite hemispheres
    ($\Delta \eta_{j_fj_f} > 4$),
      with $p_T > 15~GeV$ and
      $2.0 < |\eta_{j_f}| < 5.0$\\
(b) Forward jet invariant mass $M(j_f j_f) > 650~GeV$\\ 
(c) Narrow central jet ($|\eta_{j_\tau}| < 2$) with $p_T > 30 ~GeV$\\
(d) Central lepton ($|\eta_l| < 2$) with $p_T > 10 ~GeV$\\
(e) Lepton isolated from any other jets: $\Delta R_{lj}>0.4$

In addition, one has to take into account the $\tau$-identification 
efficiency in the one-prong channel. Here one is basically looking 
for a narrow, low-multiplicity jet whose size can be restricted, 
for example, by using the variable $R_{em}$, the `jet-radius' defined as
\cite{atlas}
\begin{eqnarray}
R_{em}~=~{\frac{\sum E_{T_i}\sqrt{(\eta_i - \eta_c)^2 + (\phi_i - \phi_c)^2}}
{\sum E_{T_i}}}
\end{eqnarray}
\noindent
where $E_{T_i}$ is the transverse energy recorded by the $i th$ cell of the 
electromagnetic calorimeter, and $i$ runs over all such cells contained 
in a cone 
of size $\Delta R~=~0.7$ (with $\Delta R^2~=~\Delta \eta^2~+~\Delta \phi^2$) 
around the jet axis, defined by ($\eta_c,\phi_c$). In addition, 
one may use the `isolation criterion' 
or the `multiplicity criterion' as defined in \cite{atlas}.
We have based our results primarily on the variable
$R_{em}$.  Thus we confine ourselves to $R_{em} < 0.07$ 
corresponding to the peak of the $R_{em}$ distributions of 
simulated $\tau$-events with $p_T$ in different ranges, 
thereby obtaining the following $\tau$-identification efficiencies in the 
hadronic channels \cite{atlas}:
\begin{eqnarray}
\epsilon_{\tau} &=& 0.30 ~for ~30 ~GeV \le p_T(j_\tau) < 50 ~GeV \nonumber \\
		 && 0.38 ~for ~50 ~GeV \le p_T(j_\tau) < 70 ~GeV \nonumber \\
		 && 0.46 ~for ~70 ~GeV \le p_T(j_\tau) \nonumber
\end{eqnarray}
The $R_{em}$ cuts also give us the factor by which non-tau jets faking 
the signal get reduced. This factor turns out to be about $400$ 
corresponding to the $\tau$-identification efficiencies listed above, 
and it has a big role in handling the backgrounds. 

The following backgrounds are found to pose the largest threat to our
suggested signals:\\
(a) $pp \longrightarrow \tau \tau  jj$ (including Drell-Yan production with QCD jets as well
as electroweak production via gauge boson fusion)\\
(b) $pp \longrightarrow W jjj$, with one jet faking the tau and
the W decaying leptonically.\\
(c) $pp \longrightarrow WW jj$, with one W decaying into a tau and the
other, into an electron or a muon. \\
(d) $pp \longrightarrow t\bar{t} X$

Although the $t\bar{t}\,+\,jets$ background looks formidable, it can still be brought under control with appropriate cuts as can be seen from table 1. We have also employed a b-veto corresponding to a b-tagging efficiency of 60\%.


For the backgrounds, we have also assumed a veto on central jets with
$p_T \le 30$ GeV, and used a veto survival probability of approximately
50\% (15\%) for colour-singlet exchange (colour exchange) processes
\cite{surv}.
After this survival probability is folded in, the $\tau\tau$  
background retains comparable contributions from electroweak and QCD 
subprocesses, the electroweak ones being mainly driven by a real Z boson.  
The $Wjjj$  background comes overwhelmingly from QCD subprocesses, 
while  $WWjj$ has mostly from electroweak contributions. Apart from 
exploiting the jet reduction factor arising out of the $R_{em}$-cut,
we also demand that the $\tau$-induced central jet and the central 
lepton have opposite electric charges, whereby the $Wjjj$ background 
gets further halved. The lepton isolation cut, imposed 
from the very beginning,  effectively suppresses backgrounds from heavy 
flavour production. We  have also found very little faking of 
the signal by pair-produced charged Higgs bosons \cite{moretti}.

In order to reduce the still remaining backgrounds, we have adopted 
the following criteria in addition to the basic cuts:\\
(a) The forward jet pair invariant mass has to be greater than 1200 GeV \\ 
(b) Missing  $E_T$ must be greater than 100 GeV \\
(c) Invariant mass of the central lepton and the tau-jet has to be greater than 60 GeV.\\ 
In table 1 we indicate how the different types of background 
as well as the signal are affected by the additional cuts. 
The signal includes contributions of comparable orders from
electroweak gauge boson fusion and real emission corrections to Drell-Yan processes,
after the central jet veto survival probabilities are folded in.
Backgrounds arising from sources other than 
gauge boson fusion undergo a drastic reduction on raising
the invariant mass cut on the forward jet pair.
Also, the strong missing-$E_T$ cut and
the invariant mass cut for the $\tau$-jet-lepton pair strongly suppress 
backgrounds.  In fact, we found by explicit analysis that 
the $b\bar{b}$ background (with two forward jets)
which can be menacing for Higgs detection  is eliminated via the missing-$E_T$ 
cut together with the demand that no jet with $E_T \ge$ 5 GeV is to be found
within a cone of $\Delta R~=~0.4$ around the central lepton. 
On the other hand, both the above cuts are 
survived with relative ease by the signal, especially when the
LSP is heavy. This immediately identifies the scenarios where
signals of the suggested type have higher chances of detection. 
\begin{figure}[h]
\includegraphics[scale=.48]{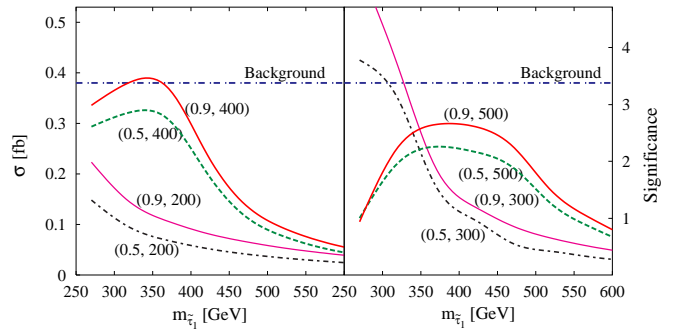}
\caption{\label{fig1} Variation of signal cross section with lighter stau mass 
in a model independent study, with $\Delta M_{\tilde\tau_1\tilde\tau_2}$ = 30 
GeV. 
The parameters {\it ($\cos \theta_\tau$, $M_2~in~GeV$)} are as shown in the 
labels. We also show 
the background cross section and significance 
($S/\sqrt{B}$). We have used $\mu $ = 500 GeV and $\tan\beta$ = 35.}
\end{figure}
In figure 1 we present our results by considering the SUSY parameter space 
in a model-independent manner, assuming the two stau mass
eigenstates to be the only `light' sfermions. The real and symmetric stau 
mass matrix is fixed in terms of its two eigenvalues and the left-right
mixing angle $\theta_\tau$. Two values of the mixing angle
have been considered, along with different values of the
SU(2) gaugino mass $M_2$ (assuming gaugino mass unification). 
The behaviour of the graphs can be traced  
to the interplay of a number of factors. First, SU(2) gauge coupling causes 
an enhancement at the production level
if the lighter stau eigenstate has a larger component of $\tilde{\tau}_L$ 
(larger $\cos\theta_{\tau}$). Secondly, a larger $\cos\theta_{\tau}$ means 
less Bino component
in the lighter stau, and therefore a suppression in its branching ratio
for decay into the LSP. Thirdly, for any value of $M_2$, higher values
of $m_{\tilde{\tau}_1}$ leads to the opening of the decay channels into
$\chi^{\pm}_1$ or $\chi^{0}_2$, and a consequent dilution of the signal.
Fourthly, as the signal level itself, there is a further $\theta_\tau$
dependence in the W/Z-induced diagrams. And finally, the signal
falls for smaller mass difference between the decaying stau and 
$\chi^{0}_1$, since  the decay leptons become too soft to
pass the  $p_T$ cuts. On the whole, however, the signal rates are
quite encouraging. In terms of $S/\sqrt{B}$ (S(B) being the number of
signal(background) events), the signal can be seen at 2-4 $\sigma$
level with an integrated luminosity of 30 $fb^{-1}$, 
for $\tilde{\tau}$-masses ranging from 250 GeV to nearly 500 GeV.
We have also checked that the lighter stau, so long as it is within
425 (450) GeV, can be detected at the $3\sigma(2\sigma)$ level even if 
the $\tilde{\tau}_2$ is much heavier.
\begin{figure}[h]
\includegraphics[scale=.60]{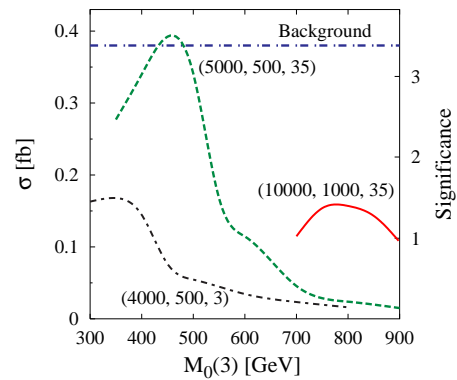}
\caption{\label{fig2} Variation of signal cross section with lighter stau mass 
for the scenario in \cite{gutcond}. The 
parameters {\it ($m_0(1)$ in GeV,$m_{1/2}$ 
in GeV, $\tan\beta$)} are as shown in the labels. We have used 
$m_0(1)=m_0(2)$ and $Sign(\mu)$ = +1.}
\end{figure}
To show the results in specific models, we present in figure 2 the 
estimated signal rates for a scenario of the kind studied in \cite{gutcond}, 
where specific boundary conditions at the GUT scale have been assumed. 
The third generation
scalar mass parameter  $m_0(3)$ is here lower than that 
corresponding to the first two, 
and consequently the two stau eigenstates emerge as the only 
non-strongly interacting sfermions in the detectable range. 
For a large gaugino mass parameter, it is not possible to go to very small
$m_0(3)$ since it will lead to a stau LSP. Thus we are restricted in 
such cases to large stau masses whose production rates are kinematically 
suppressed. A very small gaugino mass parameter, on the other hand, 
leads to problems with radiative electroweak symmetry breaking.
Thus the parameter space of this kind of a scenario is more constrained 
than in the `model-independent' cases of figure 1. Just as in the
previous case, the fall in the event rates for lower mass difference
between the stau and the LSP is due to the reduction of hardness
of the decay leptons. The region most favourable
for detection here turns out to be one where the gaugino mass  is on the 
order of 500 GeV, leading to an LSP in the mass range 200 - 250 GeV. In such
cases, particularly for large values of $\tan\beta$, one can probe
values of $m_0(3)$ up to 550 - 600 GeV at the 2$\sigma$ level. This 
corresponds to the mass of the lighter stau being upto about 500 GeV. 
The gauge boson fusion channel, therefore, appears to be the best way of 
uncovering the non-strongly interacting matter sector here.

We conclude by summarising our main observations. Supersymmetric scenarios
with inverted mass hierarchy often have the stau as the only sfermion
within the search limits of the LHC, and its mass reach 
via Drell-Yan production can be severely limited. 
It is difficult in such cases to get unambiguous
signatures of the non-strongly interacting sector of the SUSY scenario.
We have shown that the gauge boson fusion channel provides a rather
spectacular way of increasing the visibility of the superparticle spectrum
in such situations. Such visibility is at its maximum when the lighter
stau eigenstate is able to decay into only the lightest neutralino
which is sufficiently massive to carry away an appreciable amount of 
missing $p_T$. On the whole, channels of the type explored here can raise
the search limits for inverted mass hierarchy scenarios  by a 
hundred to three hundred GeV's compared to the conventional strategies.

\newcommand{\plb}[3]{{Phys. Lett.} {\bf B#1}, #2 (#3)}  
\newcommand{\nprl}[3]{Phys. Rev. Lett. {\bf #1}, #2 (#3) }
\newcommand{\prep}[3]{Phys. Rep. {\bf #1}, #2 (#3)} 
\newcommand{\rpp}[3]{Rep. Prog. Phys. {\bf #1}, #2 (#3)}
\newcommand{\nprd}[3]{Phys. Rev. {\bf D#1}, #2 (#3)}
\newcommand{\np}[3]{Nucl. Phys. {\bf B#1}, #2 (#3)}
\newcommand{\npbps}[3]{Nucl. Phys. B (Proc. Suppl.){\bf #1}, #2 (#3)}
\newcommand{\sci}[3]{Science {\bf #1}, #2 (#3)} 
\newcommand{\zp}[3]{Z.~Phys. C{\bf#1}, #2 (#3)} 
\newcommand{\epj}[3]{Eur. Phys. J. {\bf C#1}, #2 (#3)}
\newcommand{\mpla}[3]{Mod. Phys. Lett. {\bf A#1}, #2 (#3)} 
\newcommand{\ajp}[3]{{\em Am. J. Phys.\/} {\bf #1} (#3) #2}
\newcommand{\jhep}[2]{{Jour. High Energy Phys.\/} {\bf #1} (#2) }
\newcommand{\jpg}[3]{{J. Phys.\/} {\bf G#1}, #2 (#3)}
\newcommand{\astropp}[3]{Astropart. Phys. {\bf #1}, #2 (#3)}
\newcommand{\ib}[3]{{ibid.\/} {\bf #1}, #2 (#3)}
\newcommand{\app}[3]{{ Acta Phys. Polon.   B\/}{\bf #1}, #2 (#3)}
\newcommand{\nuovocim}[3]{Nuovo Cim. {\bf C#1}, #2 (#3)}
\newcommand{\yadfiz}[4]{Yad. Fiz. {\bf #1}, #2 (#3); 
                 Sov. J. Nucl.  Phys. {\bf #1} #3 (#4)]}
\newcommand{\jetp}[6]{{Zh. Eksp. Teor. Fiz.\/} {\bf #1} (#3), #2;
           {JETP } {\bf #4} (#6) #5}
\newcommand{\philt}[3]{Phil. Trans. Roy. Soc. London A {\bf #1}, #2 (#3)}
\newcommand{\hepph}[1]{hep--ph/#1}
\newcommand{\hepex}[1]{hep--ex/#1}
\newcommand{\astro}[1]{(astro--ph/#1)}

\end{document}